\magnification 1200 
\rightline{DAMTP/96-38}
\rightline{SphT/96-033}
\vskip 2cm
\centerline{\bf A Renormalisation Group Analysis}
\vskip .2 cm
\centerline{\bf of}
\vskip 0.2 cm
\centerline{\bf 2d Freely Decaying  Magnetohydrodynamic Turbulence}
\vskip 1 cm
\centerline{by}
\vskip .2 cm
\centerline{Ph. Brax\footnote{$^*$}{On leave of absence from SPhT-Saclay
CEA F-91191 Gif sur Yvette}}
\vskip .5 cm
\centerline{DAMTP}
\centerline{University of Cambridge}
\centerline{Silver Street}
\centerline{Cambridge CB39EW UK}
\vskip 3 cm
\leftline{\bf Abstract:}
\vskip .2 cm
We study two dimensional
 freely decaying magnetohydrodynamic turbulence.
 We
investigate 
 the time evolution of the probability law of the gauge field and the
stream function. 
Assuming that this probability law is initially 
defined by a statistical field
theory in the basin of attraction of a renormalisation group fixed point,
we show that its time evolution is generated by renormalisation
transformations. In the long time regime, the probability law is
described   by
 non-unitary conformal field theories. In that case,  we prove  that
the  kinetic and magnetic energy spectra 
 are proportional. We then construct a family of fixed points 
using the $(p,p+2)$ non-unitary minimal models of conformal field
theories.

Pacs numbers: 47.27.Ak, 11.10.Gh, 05.40.+j
\vfill\eject

 It has been recently noticed$^{[1]}$ by Polyakov 
 that field theoretic methods can
be used  to study steady states of two dimensional turbulence.
 In this approach, Polyakov assumes
that the velocity field of a turbulent fluid can be described by a
primary field deduced from a conformal field theory.   
The same method  has been applied$^{[2]}$ by Ferretti and Yang in the case of 
steady states of 
magnetohydrodynamic turbulence.  
Other possible applications to magnetohydrodynamic have been also
discussed by Coceal and Thomas$^{[3]}$.
One of the results found by 
Ferretti and Yang is that some steady states 
 do not respect the equipartition $^{[4]}$ between the
magnetic and the kinetic energies. It has been argued$^{[5]}$ by
Rahimi-Tabar and Rouhani that
 the equipartition can be satisfied provided that time as well as
spatial scale invariances hold.  
In this communication,
we shall use renormalisation group techniques to study the long time behaviour
 of freely decaying magnetohydrodynamic turbulence  in two dimensions.
In particular, we shall see that conformal field theories appear as
fixed points of the time evolution of the probability law of the stream
function and the gauge field in the long time regime. At these fixed
points, the kinetic and the magnetic energy
spectra
are proportional.
 Finally, we shall construct a family of fixed points using the
$(p,p+2)$ non-unitary minimal models.

In two dimensions, the velocity field of an incompressible fluid  is
determined by a single pseudo-scalar field: the stream function
$$v_{\alpha}=e_{\alpha\beta}\partial_{\beta} \psi.\eqno(1)$$
The tensor $e_{\alpha \beta}$ is the antisymmetric tensor $e_{12}=-e_{21}=1$.
The vorticity is then given by
$$\omega=e_{\alpha \beta}\partial_{\alpha}v_{\beta}.\eqno(2)$$
It is a pseudo-scalar too.
We are  interested in the two dimensional situation where the
magnetic field lies in the plane of the fluid motion. Moreover we
suppose that  there
is no external magnetic field and the magnitude of the initial magnetic field
is of the same order as the magnitude of the initial velocity field (in
well-chosen units). This situation can occur 
 for certain astrophysical phenomena
where the magnetic Reynolds number is large.  
In the absence of magnetic monopoles, the magnetic field can also be
represented by
$$B_{\alpha}=e_{\alpha\beta}\partial_{\beta} A,\eqno(3)$$
where $A$ is a scalar gauge field. The induced electric current can be
deduced 
$$J=e_{\alpha\beta}\partial_{\alpha}B_{\beta}.\eqno(4)$$
These quantities satisfy coupled partial differential equations.
 We will focus on the free decaying
case in an infinite plane. We shall suppose that the vorticity field
and the electric current decay sufficiently fast at infinity.

The Navier-Stokes equations are effective equations describing the
behaviour of a fluid in the hydrodynamic approximation. 
The velocity field is the average of the microscopic velocity of
molecules over a macroscopic size $a$, the size of a fluid element.
  Similarly, the magnetic field is also 
 a macroscopic
field.  We shall  analyse  the macroscopic equations for
the 
regularised vorticity and gauge fields$^{[2]}$
$$\eqalign{
\partial_t \omega_a+
e_{\alpha\beta}\partial_{\alpha}\psi_a\partial_{\beta}\Delta
\psi_a&=e_{\alpha\beta}\partial_{\alpha}J_a\partial_{\beta}A_a+\nu
\Delta \omega_a\cr
\partial_t A_a +v_{a\alpha}\partial_{\alpha}A_a&=\eta J_a.\cr}
\eqno(5)$$
The viscosity $\nu$ and the resistivity $\eta$ are responsible for
dissipation.  
As the resistivity and the viscosity have the same dimensions (in
well-chosen units) , one deduces
$$\eqalign{
\nu&={a^2\over \tau}\cr
\eta&= \eta_0 {a^2\over \tau},\cr}
\eqno(6)$$
where $\eta_{0} $ is of order one. The time $\tau$ is a characteristic
time for the decay of both the kinetic and the magnetic energies.

In field theoretic terms, the
size $a$ plays the role of an intrinsic ultraviolet cut-off. We shall
be interested in the situation where
 the gauge field and the stream function 
are  random fields regularised at short distances by the molecular
size   $a$. 
 As no  random stirring is imposed,
 we shall suppose that initial conditions are randomly
prescribed by a probability law for both the initial magnetic and
velocity fields. The evolution of a given configuration is given by
(5). The evolution of the probability law of the gauge field and
the stream function $d{\cal P}(A_a,\psi_a)$
 is governed by the Hopf equations obtained
considering (5) as valid in any equal time correlation functions of the gauge
field and the stream function
$$\eqalign{
\ &{\partial\over \partial
t}<\omega_a(x_1,t)...\omega_a(x_n,t)A_a(x_{n+1},t)...
A_a(x_{n+m},t)>\cr
\ &\sum_{i=1}^n <\omega_a(x_1,t)...(
-e_{\alpha\beta}\partial_{\alpha}\psi_a(x_i,t)\partial_{\beta}\Delta
\psi_a(x_i,t)+e_{\alpha\beta}\partial_{\alpha}J_a(x_i,t)\partial_{\beta}
A_a(x_i,t)\cr
\ & +\nu
\Delta \omega_a(x_i,t))...\omega_a(x_n,t)A_a(x_{n+1},t)...
A_a(x_{n+m},t)>+\cr
&\sum_{i=n+1}^{n+m}<\omega_a(x_1,t)..\omega_a(x_n,t)A_a(x_{n+1},t)..
(-v_{a\alpha}(x_i,t)\partial_{\alpha}A_a(x_i,t)+\eta
J_a(x_i,t))..A_a(x_{n+m},t)>.\cr}
\eqno(7)$$
At each time, the probability law for both the magnetic and velocity
fields is modified. We shall suppose that this probability
law at time $t$ after a transient period of $O(\tau)$ 
becomes   a statistical field theory described   by a
Boltzmann weight $\exp(-S_{t,\tau})$ 
$$d{\cal P}(A_a,\psi_a)={1\over Z}dA d\psi
\exp(-S_{t,\tau}(A_a,\psi_a))\eqno(8)$$
where the partition function $Z$ is a normalising factor. The action
$S_{t,\tau}$ is a function of $t\over \tau$.
 We shall be
interested in the subsequent evolution of this probability law.

 The equations (7) can be shown to be
invariant under renormalisation transformations. Let us denote by $d_A$ the
scaling dimension of $A$ and $d_{\psi}$ the scaling dimension of
$\psi$ obtained from the field theory (8). These scaling dimensions
are well-defined when the field theory (8) is in the basin of
attraction of a fixed point of the renormalisation group. In that
case, the dimensions $d_A$ and $d_{\psi}$ are the conformal dimension
of the field $A$ and $\psi$ at the fixed point. Let us analyse the
solutions of (7). 
 The evolution of solutions of (7) is given by
$$\eqalign{
\tilde A_a(\lambda x, \lambda^T t)=\lambda^{-d_A}A_{\lambda^{-1}a}(x,t)\cr
\tilde\psi_a(\lambda x, \lambda^T
t)=\lambda^{-d_{\psi}}\psi_{\lambda^{-1}a}(x,t)\cr} 
\eqno(9)$$
and for the action 
$$e^{ -\tilde S_{\lambda^T t,\tau}
(\tilde A_a(\lambda x,\lambda^T t),\tilde\psi_a(\lambda
x,\lambda^T t))}=\int_{[{a\over\lambda},a]}[dA][d\psi]
e^{ -S_{t,{\tau\over \lambda^T}} 
(A_{\lambda^{-1}a}(x,t), \psi_{\lambda^{-1}a}(x,t))}\eqno(10)$$
where $\tilde S_{\lambda^T t,\tau}$ is the action at time $\lambda^T t$ 
for the solutions 
 defined
by (9). The right-hand side of (10) denotes a renormalisation group
transformation where one integrates over all fluctuations of $A$ and
$\psi$ in the range $[{a\over\lambda},a]$. This renormalisation group
transformation is carried out in momentum space where modes in the
range $[a^{-1},\lambda a^{-1}]$ are integrated out. We see
that the probability law of the stream function remains a statistical
field theory given by the renormalised Boltzmann weight (10).

Let us show that (9) and (10) give the evolution of 
the probability law of the gauge
field and the stream function.
Let us first write (5) at time $t$ after transforming $a\to
\lambda^{-1}a$
and $\tau\to \lambda^{-T}\tau$. This reads
$$\eqalign{
\partial_t \omega_{\lambda^{-1}a}+e_{ \alpha
\beta}\partial_{\alpha}\psi_{\lambda^{-1}a}\partial_{\beta}\Delta
\psi_{\lambda^{-1}a}&=\nu\lambda^{T-2}\Delta
\omega_{\lambda^{-1}a}+e_{\alpha
\beta}\partial_{\alpha}J_{\lambda^{-1}a}\partial_{\beta}A_{\lambda^{-1}a}\cr
\partial_{t}A_{\lambda^{-1}a}+v_{\lambda^{-1}a,\alpha}\partial_{\alpha}
A_{\lambda^{-1}a}&=\eta\lambda^{T-2}J_{\lambda^{-1}a}.\cr}
\eqno(11)$$
These equations are valid at time $t$ in any correlation functions of the
gauge field and the stream function. Using (9), one can substitute
 $\lambda^{d_A}A_a$ and $\lambda^{d_{\psi}}\psi_a$ for
$ A_{\lambda^{-1}a}$ and $ \psi_{\lambda^{-1}a}$. In any correlation
functions, the functional integral over modes in $[a^{-1},\lambda
a^{-1}]$ can be carried out and involves only the Boltzmann
weight. The integration of the Boltzmann weight gives (10). One then
obtain  the equations
$$\eqalign{
\partial_{\lambda^T t} \omega_a+
\lambda^{2+d_{\psi}-T} 
e_{\alpha\beta}\partial_{\lambda x_{\alpha}}
\psi_a \partial_{\lambda x_{\beta}}\partial^2_{\lambda x}
\psi_a&=\lambda^{2+2d_A-d_{\psi}-T} 
e_{\alpha\beta} \partial_{\lambda x_{\alpha}}
J_a \partial_{\lambda x_{\beta}}A_a+\nu
\partial^2_{\lambda x} \omega_a\cr
\partial_{\lambda^T}t A_a +\lambda^{2+d_{\psi}-T}v_{a\alpha}
 \partial_{\lambda x_{\alpha}}A_a&=\eta J_a.\cr}
\eqno(12)$$
These equalities are satisfied in any correlation functions at time
$\lambda^T t $ and coordinates $\lambda x$. They coincide with (5)
provided the time rescaling exponent and the dimensions of the fields satisfy
$$\eqalign{
T&=d_{\psi}+2\cr
d_A&=d_{\psi}.\cr}
\eqno(13)$$
These conditions guarantee that the evolution  of the solutions of (7)
is given by (9) and (10).
 The time evolution of solutions
is then  well-understood when the initial probability law of the gauge field
and the stream function (8) is in the basin of attraction of a fixed
point of the renormalisation group. In that case, the probability law
of the gauge field and the stream function converges to the
probability law defined by the fixed point in the long time
regime. Moreover the fixed point is a conformal field theory
$^{[6]}$. 

Let us analyse the physical consequences of (13). 
 The 2-point
correlation functions of the magnetic field and the velocity field is
characterised by the scaling dimension (conformal dimension)
$d_A=d_{\psi}$.
It is easily seen that in the long time regime, the energy spectra
converge to
$$\eqalign{
E_{\hbox{mag}}(k)\sim k^{2d_{A}+1}\cr
E_{\hbox{kin}}(k)\sim k^{2d_{\psi}+1}\cr}
\eqno(14)$$
in the  range $k\ll a^{-1}$.
 The equality between the scaling
dimensions of the gauge field and the stream function implies that
these two spectra are proportional. This result characterises the
extreme correlation between the velocity field and the stream
function.
In the following, we shall further analyse the possible fixed points
and determine examples of energy spectra.

 In the long time regime, solutions describe a turbulent state if the
 viscosity and the resistivity become
negligible.  This implies that the dimension of $\psi$ has to be
negative in order to ensure that the rescaled viscosity (respectively
resistivity) 
$\nu(\lambda)=\lambda^{d\psi} \nu$ (respectively
$\eta(\lambda)=\lambda^{d_{\psi}} \eta$) deduced from (11) converge to zero.
The fixed points are therefore non-unitary conformal field theories.
The exponent $T$ being positive, one deduces
$$-2<d_{\psi}<0.\eqno(15)$$
The second inequality implies that the dimensions of the electric
current $J$ and the vorticity $\omega$ are positive. This guarantees
that the current $J$ and the vorticity $\omega$ decrease sufficiently
rapidly at infinity.

 Solutions can asymptotically reach a time independent non-unitary 
fixed point
if and only if all the non-linear terms in (11) converge to zero. The
non-linear terms can be readily evaluated when the cut-off is rescaled
to zero using properties of short distance expansions. 
Following Polyakov, we shall suppose that the gauge field and the stream
function become primary fields at the fixed point with conformal
dimensions $d_A$. Then the short distance expansion of these fields
can be summarised using conformal families
 ($[A]$ is the conformal family of $A$, i.e. $A$ and all its
descendent fields)
$$\eqalign{
[A][A]=[A_2]+...\cr
[\psi][\psi]=[\psi_2]+...\cr
[A][\psi]=[\chi]+...,\cr}
\eqno(16)$$       
where dots mean higher order terms. Therefore the non-linear terms read
$$\eqalign{
e_{\alpha\beta}\partial_{\alpha}\psi_{\lambda^{-1}a}\partial_{\beta}\Delta
\psi_{\lambda^{-1}a}&\sim
({a\over\lambda})^{d_{\psi_2}-2d_{\psi}}(L_{-2}\bar L_{-1}^2-\bar
L_{-2}L_{-1}^2 )\psi_{\lambda^{-1}a,2}+...\cr
e_{\alpha\beta}\partial_{\alpha}J_{\lambda^{-1}a}\partial_{\beta}
A_{\lambda^{-1}a}&\sim ({a\over\lambda})^{d_{A_2}-2d_A}(L_{-2}\bar L_{-1}^2-
\bar L_{-2} L_{-1}^2)A_{\lambda^{-1}a,2}+...\cr
v_{\alpha, \lambda^{-1}a}\partial_{\alpha}A_{\lambda^{-1}a}&\sim 
({a\over\lambda})^{d_{\chi}-d_A-d_{\psi}+2}(L_{-2}\bar L_{-1}^2- \bar
L_{-2} L_{-1}^2)\chi_{\lambda^{-1}a}+...,\cr}
\eqno(17)$$
where  $L_n={1\over 2\pi i}\int
dz\ z^{n+1} T(z)$ are the Virasoro generators deduced from the energy
momentum tensor $T$ of the fixed point.
Using the equality between the dimensions of
the stream function and the gauge field, these products vanish
asymptotically if
$$\eqalign{
d_{\psi_2}&>2d_{\psi}\cr
d_{A_2}&>2d_{A}\cr
d_{\chi}&>2d_{\psi}-2.\cr}
\eqno(18)$$
Any non-unitary conformal field theory satisfying these criteria is a
time-independent fixed point of the magnetohydrodynamics
equations\footnote{$^1$}{The -2 was erroneously missing in
Ref.[3]$^{[6]}$}.

In order to get more information on the fixed points, we shall
 be interested in the fluxes of the energy $E={1\over 2}\int d^2
x(v_a^2+B_a^2)$
and the magnetic enstrophies
$M_n=\int d^2 x A_a^n$.
In the Kolmogorov picture$^{[8]}$
 of turbulent steady states, these fluxes are
constant. In particular, the constraint used by Ferretti and Yang
concerns the magnetic enstrophy $(n=2)$. In the decaying  case, one
can expect these fluxes to vanish in the long time regime as no
forcing is imposed.   
 The  fluxes associated to  
 the magnetic enstrophies and the energy are $^{[1,2]}$ 
$$\eqalign{
R_{n,a}(q)&=-n\int d^2x\ \theta_q*(v_{a\alpha}\partial_{\alpha}A_a\
 A_a^{n-1}(x)) \cr
\epsilon_a(q)&=-  \int d^2x \theta_q*(e_{\alpha\beta}
\partial_{\alpha}\psi_a\partial_{\beta}\Delta \psi_a\psi_a),\cr}
\eqno (19)$$
where $\theta_q$  is an approximate delta function
$\theta_q(x)=\int_{k>q}d^2k \ e^{2\pi i k.x}.$
The fluxes $R_n$ and $\epsilon$ 
represent the transfer rate of magnetic enstrophies and energy 
in momentum space.    
In the asymptotic regime, the $q$ dependence of $R_{n,a}$
disappears. Defining  the product of n copies of $A$ with
itself, the product of $n$ copies of $\psi$ 
and the product of n copies of $A$ with $\psi$ $^{[6]}$
$$\eqalign{
[A]...[A]=[A_{n}]+...\ \cr
[\psi]...[\psi]=[\psi_n]+...\ \cr
[\psi][A]...[A]=[\chi_{n+1}]+...\cr}
\eqno(20)
$$
where dots denote higher order terms, the fluxes behave in the long
time regime as 
$$\eqalign{
R_{n,a}&=a^{2+d_{\chi_{n+1}}-(n+1)d_{\psi}}\int d^2x
(L_{-2}\bar L_{-1}^2-\bar L_{-2}L^2_{-1})
\chi_{n+1,\lambda^{-1}a}\cr
\epsilon_a&=a^{d_{\psi_3}-3d_{\psi}}\int d^2x
(L_{-2}\bar L_{-1}^2-\bar L_{-2}L^2_{-1})\psi_{3,\lambda^{-1}a}.\cr}
\eqno(21)$$
The appearance of a descendent field of $\chi_{n+1}$ on the right hand
side of (21) stems from the properties of the field 
$ \chi_{n+1}$  under parity (similarly for $\psi_3$).
 Notice that $r_{n,a}$ is a scalar and $\chi_{n+1}$
is a pseudo-scalar as $\psi$ is a pseudo-scalar. 
 The right hand side of (21) is invariant under parity
as it is constructed using a combination of Virasoro generators 
which picks up a minus sign under a parity transformation.
 This combination is the one we had already encountered
when calculating the non linear terms of the equations (5).
The result (21) is valid provided 
$$\eqalign{
d_{\chi_{n+1}}&=d_{A_{n-1}}+d_{\chi}\cr
d_{\psi_3}&=d_{\psi}+d_{\psi_2}.\cr}
\eqno(22)$$
In the  decaying case, fluxes given by (21) have vanishing expectation
values.
Moreover, the fluxes vanish at the fixed points
$$\eqalign{
R_n&=0\cr
\epsilon&=0\cr}
\eqno(23)$$
when  the dimensions of 
$(L_{-2}\bar L_{-1}^2-\bar L_{-2}L^2_{-1})
\chi_{n+1}$ and $(L_{-2}\bar L_{-1}^2-\bar L_{-2}L^2_{-1})\psi_{3}$
 are  different from two. This is expected as there is no
forcing on large scales.

In general the constraints (22) are difficult to analyse. We shall
 construct a particular family of non-trivial fixed points of 2d
freely decaying magnetohydrodynamic turbulence. The simplest examples
of non-unitary conformal fields can be drawn from minimal models
$^{[6]}$.
These models possess a finite number of primary fields. 
In order to distinguish scalars from pseudo-scalars, it is convenient
to introduce a parity factor equal to one for scalars and -1 for
pseudoscalars. 
 Let us suppose
that the  gauge field $A$ becomes equal to the 
dual of the stream function at the
fixed point, i.e. the scalar field $A=A\otimes 1$  differs from
the pseudo-scalar $\psi=A\otimes -1$ only by the parity factor.
 This is compatible with
the equality $d_A=d_{\psi}$.   
Let us consider now  the $(p,p+2)$ non-unitary minimal
models ($p>1$ and odd). One can identify $A$ with the  unique
field  of negative dimension  such
that $[A][A]=[I]+...\ $. 
In particular, one gets
$$d_A=-{3\over 2p(p+2)}.\eqno(23)$$
 This implies that $A_2=I$, $\psi_2=I$ and
$\chi=\tilde I$ where $\tilde I$ is the dual of the identity $I$. The
conditions (18) are automatically satisfied. Let us now check the
requirements (22). Using the orthogonality between primary fields, one
can deduce that $[A][A][A]=[A]$. This entails that $A_n=I$ for $n$
even and $A_n=A$ for $n$ odd. Similarly one obtains $\psi_3=\psi$, 
$\chi_{n+1}=\tilde I$ for
$n$ odd, and $\chi_{n+1}=\psi$ for $n$ even. This implies that (22) is
identically satisfied. This proves that $(p,p+2)$ minimal models
provide a family of fixed points of 2d magnetohydrodynamics. It is
worth noting that in this case the magnetic and the kinetic 
energy spectra are equal
$$E(k)\sim k^{-{3\over p(p+2)}+1}.\eqno(25)$$
This spectrum is always integrable.
The $(p,p+2)$ minimal models describe finite energy density solutions
of 2d freely decaying magnetohydrodynamic turbulence in the long time regime.

Let us comment on the similarities between the decaying case and
steady states of 2d magnetohydrodynamic turbulence. 
 In both situations, conformal
field theories are used to describe the probability law of the gauge
field and the stream 
function. For steady states, conformal invariance is an assumption
used in order to describe the scale invariant regime in the inertial
range. In the freely decaying case, one can control the time evolution
of solutions and prove that conformal field theories appear in the
long time regime. In both problems, the vanishing of (17) follows from
scale invariance in a regime where
the viscosity and the resistivity are
 negligible. Another
similarity is the appearance of non-unitary theories. For steady
states, this is a consequence of the constant magnetic enstrophy flux
condition. In the free decaying case, the non-unitarity comes from the
requirement that the renormalised viscosity $\nu
(\lambda)=\lambda^{d_{\psi}}\nu$ and the renormalised resistivity 
$\eta(\lambda)=\eta \lambda^{d_{\psi}}$ become
 negligible in the long time regime. In
both cases, the fact that the viscosity is non-zero plays a
fundamental role. This reinforces Polyakov's idea$^{[1]}$ that non-unitary
theories should describe turbulence as turbulence  is  
a flux state where
dissipation takes place on  small scales and not an equilibrium
state.

 Finally, there are some relevant differences between steady
states and the freely decaying case. First of all, solutions of 2d
freely decaying magnetohydrodynamic turbulence do not require parity
violation whereas it is easier to understand steady states of 2d
magnetohydrodynamic turbulence if parity is violated. We
have 
also found  that $d_{\psi}>-2$ at a fixed point.
 This condition ensures that the
vorticity and the electric current decay at infinity.
 Obviously, this is intimately linked to
the absence of boundaries. It is certainly a different situation for
steady states where boundary effects play a significant role. The same
condition can be written $T>0$, it is then connected to the breaking
of time reversal invariance. Indeed, this constraint implies that
solutions at later times $\lambda^T t$ are obtained after averaging
over small scale features at time $t$. Therefore in the process of
evolution, small scale details of the gauge field and the 
stream function are forgotten. 
This implies that one cannot deduce  solutions at time $t$ from
solutions at time $\lambda^T t$. In that sense, the behaviour of
solutions of magnetohydrodynamics  is
irreversible. Eventually, 
 the equipartition of magnetic and kinetic energies is not
automatically satisfied for steady states.  
We have seen that the
 kinetic and the magnetic energy spectra are proportional
 in the asymptotic regime of
the decaying case. Unfortunately, our approach does not allow to
determine the proportionality constant relating the kinetic and
magnetic energies.

\vskip .5 cm
In conclusion, we  have studied  the long time evolution of 2d freely decaying
magnetohydrodynamic turbulence assuming that the initial 
probability law of the
gauge field and the stream function can be described by a statistical
field theory in the basin of attraction of a renormalisation group
fixed point. We have  shown that the time evolution of this
probability law is given by renormalisation trajectories. In that case,
the long time regime is determined by non-unitary fixed points of the
renormalistion group. 
In the asymptotic regime, the kinetic and the magnetic energy spectra
are proportional. 
 We have then constructed a family of fixed points
using non-unitary minimal models of conformal field theories. The
existence of a large number of fixed points provides an
explicit example of non-universality. The convergence towards a
particular fixed point is indeed dependent on the initial conditions.
It would be extremely interesting to understand the dynamics of 2d
magnetohydrodynamic turbulence when the initial conditions are
arbitrary. In particular, the influence of the infinite number of
fixed points on the trajectories of solutions is a noteworthy problem.

Acknowledgements:
I am grateful to W. Eholzer, G. Ferretti, M. Gaberdiel and
R. Peschanski for many discussions and suggestions.
\vfill\eject   
\centerline{\bf References}
\vskip 1 cm
\leftline{[1] A. M. Polyakov Nucl. Phys. {\bf B396}, 397 (1993)}
\vskip .2 cm
\leftline{[2] G. Ferretti and Z. Yang Europhys. Lett. {\bf 22}, 639
(1993)}
\vskip .2 cm
\leftline{[3] O. Coceal and S. Thomas `` Conformal Models of MHD
Turbulence''
QMW-PH 95-45 preprint}
\vskip .2 cm
\leftline{[4] S. Chandrasekahar Annals of Physics {\bf 2}, 615 (1957)}
\vskip .2 cm
\leftline{[5] M. R . Rahimi Tabar and S. Rouhani ``Turbulent 2D-
magnetohydrodynamics}
\vskip .2 cm
\leftline{and Conformal Field Theory, hep-th/9503005}
\vskip .2 cm
\leftline{[6] A. A. Belavin, A. M. Polyakov and A. B. Zamolodchikov
Nucl. Phys. {\bf B241} (1984) 333}
\vskip .2 cm
\leftline{[7] G. Ferretti Private Communication}
\vskip .2 cm
\leftline{[8] A. N. Kolmogorov: C. R. Aca. Sci. USS {\bf 243} (1941) 301.}
\end